\documentclass[aps,prc,preprint,groupedaddress,preprintnumbers,12pt]{revtex4-1}
\usepackage{natbib}
\usepackage{amsmath}
\usepackage{graphicx}
\usepackage{hyperref}
\usepackage{slashbox}
\usepackage{longtable}
\usepackage{bm}

\begin{document}

\preprint{JLAB-THY-13-1749}

\title{Off-shell extrapolation of Regge-model NN scattering amplitudes describing final state interactions in $^2H(e,e'p)$}

\author{William P. Ford$^{(1)}$}
\email[]{wpford@jlab.org}
\author{J. W. Van Orden$^{(1,2)}$}
\email[]{vanorden@jlab.org}
\affiliation{\small \sl (1) Department of Physics, Old Dominion University, Norfolk, VA
23529 \\ and\\ (2) Jefferson Lab\footnote{Notice: Authored by Jefferson Science Associates, LLC under U.S. DOE Contract No. DE-AC05-06OR23177.
The U.S. Government retains a non-exclusive, paid-up, irrevocable, world-wide license to publish or reproduce this manuscript for U.S. Government purposes},
12000 Jefferson Avenue, Newport News, VA 23606}
\date{\today}
\begin{abstract}
In this work, an off-shell extrapolation is proposed for the Regge-model $NN$ amplitudes of \cite{FVO_Reggemodel}. A prescription for extrapolating these amplitudes for one nucleon off-shell in the initial state are given. Application of these amplitudes to calculations of deuteron electrodisintegration are presented and compared to the limited available precision data in the kinematical region covered by the Regge model. 
\end{abstract}
\maketitle

\section{Introduction}

\begin{figure}
    \centerline{\includegraphics[height=3in]{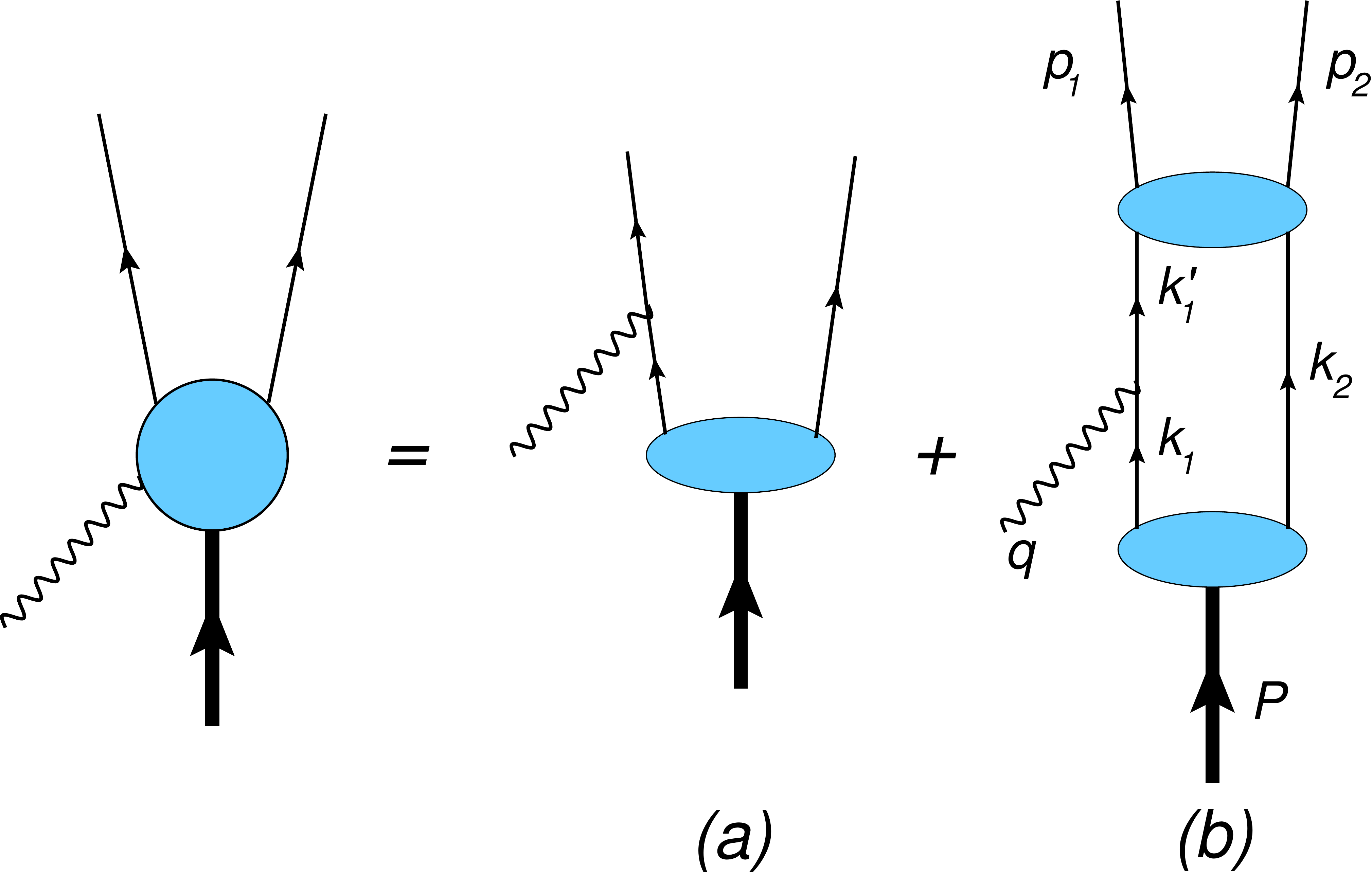}}
    \caption{(color online) Diagrams representing deuteron electrodisitegrqation at large energy and momentum transfers. Diagram (a) is the plane-wave contribution and diagram (b) includes final-state interactions (FSI). }
    \label{fig:deepia}
\end{figure}

\begin{figure}
    \centerline{\includegraphics[height=3in]{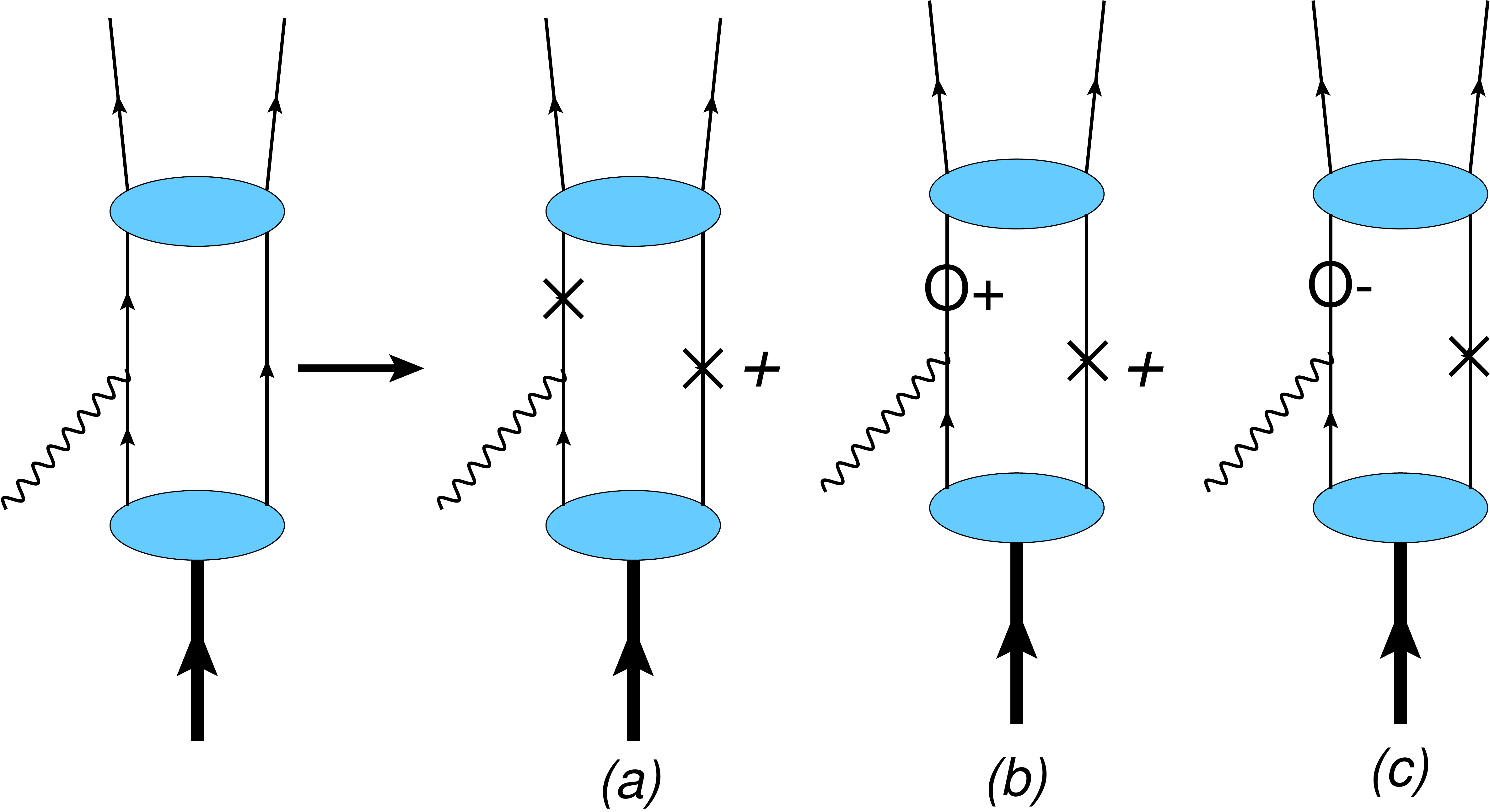}}
    \caption{(color online) Diagrams representing the separation of the FSI contribution to deuteron electrodisintegration into (a) on-shell, (b) positive energy off-shell and (c) negative energy off-shell contributions. Diagram (c) gives a small contribution and is neglected in the calculations presented here. }
    \label{fig:deepiaoff}
\end{figure}

In a recent paper \cite{FVO_Reggemodel} we described a fit to nucleon-nucleon scattering for Mandelstam $s=5.4\ {\rm GeV^2}$ to $s=4000\ {\rm GeV^2}$ using a Regge model. The immediate purpose of this model was to allow extension of calculations of deuteron electrodisintegration to higher invariant masses than was possible using the SAID helicity amplitudes as used in \cite{JVO_2008_newcalc,JVO_2009_tar_pol,JVO_2009_ejec_pol}. In \cite{JVO_2008_newcalc} the electrodisintegration amplitude for the $d(e,e'p)$ amplitude for large energy and momentum transfers was described by the Feynman diagram represented in Fig. {\ref{fig:deepia}}. Diagram Fig. {\ref{fig:deepia}}(a) represents the plane-wave (PWIA) contribution, while the diagram represented by Fig. {\ref{fig:deepia}}(b) includes the final-state interaction (FSI).
It was shown that the FSI contribution Fig. {\ref{fig:deepia}}(b) could be represented by the diagrams of Fig. {\ref{fig:deepiaoff}}. Examination of the poles in the loop integral shows that the integral is dominated by the pole in the propagator for particle 2 which is represented by the cross on this line in diagrams Fig. {\ref{fig:deepiaoff}}(a-c). The propagator for $k'_1$ can be separated into an on-shell part Fig. {\ref{fig:deepiaoff}}(a), an off-shell part with positive energy projection Fig. {\ref{fig:deepiaoff}}(b) and an off-shell part with negative energy projection Fig. {\ref{fig:deepiaoff}}(c).

In \cite{FJVO} we presented a comparison of the FSI contributions to the $d(e,e'p)$ reaction using the SAID and Regge parameterizations of the FSI. Since both the SAID and Regge helicity amplitudes are fit to on-shell data, only the on-shell contribution to the FSI represented by Fig. {\ref{fig:deepiaoff}}(a) was included. In order to understand the uncertainty in these calculations, it is necessary to have some reasonable extrapolation of the $NN$ scattering amplitudes off-shell.

The object of this paper is to provide a reasonable extrapolation of the Regge-model amplitudes for particle 1 off-shell. In section II we will show how this extrapolation is constructed and show the effects of an off-shell extrapolation of the $NN$ differential cross section. In section III we will apply this off-shell extrapolation to the $d(e,e'p)$ reaction and show its cutoff dependence. Section IV will contain a summary and conclusions drawn from this work.

\section{Off-shell extrapolation of the Regge-model NN amplitudes}

\begin{figure}
    \centerline{\includegraphics[height=3in]{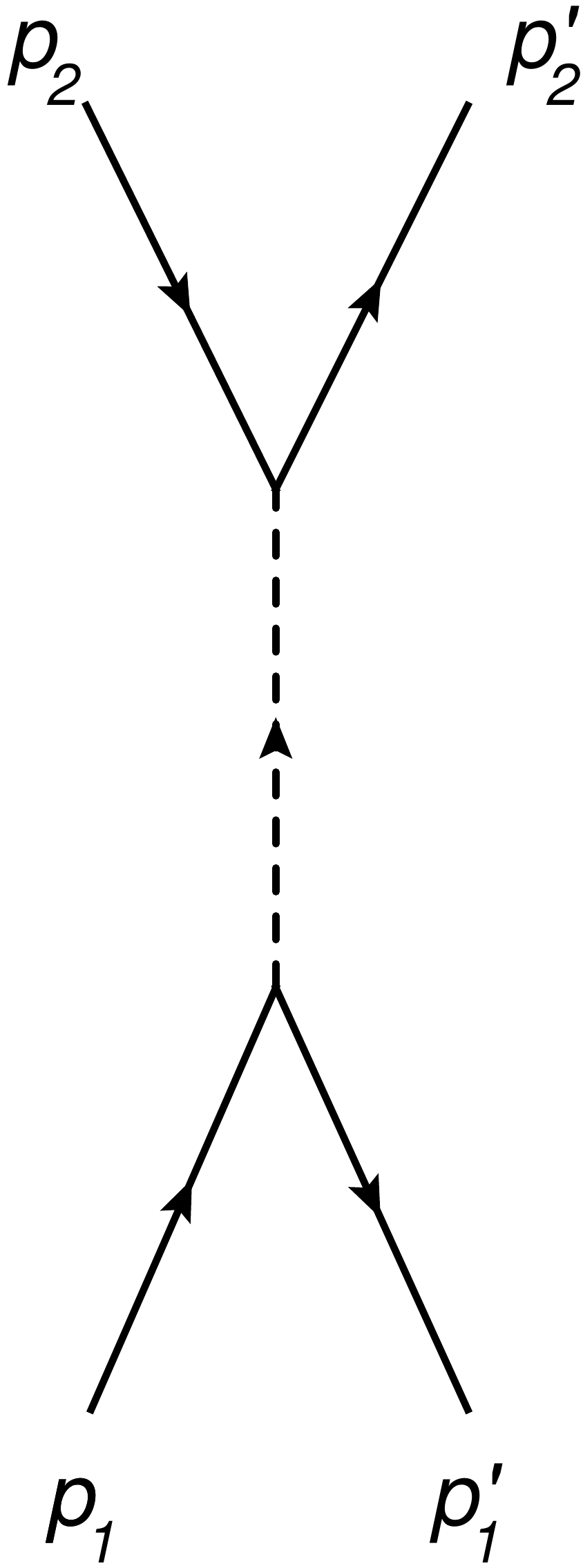}}
    \caption{(color online) Representation of the direct contribution to $N\bar{N}$ scattering through a single meson exchange in $t$-channel center of momentum frame.}
    \label{fig:tcm}
\end{figure}

As described in \cite{FVO_Reggemodel} and \cite{FJVO}, the scattering amplitudes in the $s$ channel are described by Reggion exchange in the $t$ channel. In the $t$-channel center-of-momentum (cm) frame the amplitudes are given by $N\bar{N}$ scattering as represented by Fig. {\ref{fig:tcm}}. For $p_1$ off-shell and all other legs on shell, the four-momenta are given by 
\begin{align}
p_1&=(p_1^0,\bm{p})\nonumber\\
p'_1&=(E_t,-\bm{p})\nonumber\\
p_2&=(E'_t,-\bm{p}')\nonumber\\
p'_2&=(E'_t,\bm{p}')
\end{align}
where $E'_t$ is the on-shell energy of the final-state particles, $\pm\bm{p}'$ are the momenta of the final state particles, $E_t$ is the on-shell energy for initial state particles of momentum $\pm\bm{p}$ and $p_1^0$ is the off-shell energy of particle 1. Using energy conservation, it can be easily shown that 
\begin{align}
E'_t&=\frac{\sqrt{t}}{2}\nonumber\\
|\bm{p}'|&=\sqrt{\frac{t}{4}-m^2}\nonumber\\
E_t&=\frac{t-v}{2\sqrt{t}}\nonumber\\
|\bm{p}|&=\sqrt{\frac{(t-v)^2}{4t}-m^2}
\end{align}
where
\begin{equation}
v=p_1^2-m^2
\end{equation}
is a measure of the off-shellness of particle 1 and $m$ is the nucleon mass. The scattering angle in the $t$ cm frame is
\begin{equation}
z=\cos \theta_t=\frac{2s+t-4m^2-v}{\sqrt{\left(4m^2-t\right)\left(4m^2-\frac{(t-v)^2}{t}\right)}}\,.\label{eq:zoff}
\end{equation}
For this situation the constraint on the Mandelstam variables $s$, $t$ and $u$ is given by
\begin{equation}
s+t+u=3m^2+p_1^2=4m^2+v\,.
\end{equation}

In \cite{FVO_Reggemodel} we used helicity matrix elements of the Fermi invariant representation to relate the scalar functions in the Fermi invariant representation to the Reggeized matrix elements in the $t$ cm frame. This approach greatly simplifies the extrapolation of the scattering amplitudes to the physical $s$ cm frame. For on-shell scattering there are five terms in the Fermi invariant description of the scattering matrix characterized by the scalar functions ${\cal F}_{S}$, ${\cal F}_V$, ${\cal F}_T$, ${\cal F}_P$ and ${\cal F}_A$. If $p_1$ is off-shell, there are five additional scalar functions ${\cal F}_{SO}$, ${\cal F}_{VO}$, ${\cal F}_{T0}$, ${\cal F}_{PO}$ and ${\cal F}_{A0}$. The Fermi-invariant representation of the off-shell scattering operator can then be written as
\begin{align}
   \hat{M} &= {{\cal F}_{S}(s,t,v)}1^{(1)}  1^{(2)}+ {{\cal F}_V(s,t,v)} \gamma^{\mu(1)} \gamma_{\mu}^{(2)}
           + {{\cal F}_T(s,t,v)} \sigma^{\mu \nu (1)} \sigma_{\mu \nu}^{(2)} \nonumber \\
           &- {{\cal F}_P(s,t,v)} (i\gamma_5)^{(1)}  (i\gamma_5)^{(2)}
           + {{\cal F}_A(s,t,v)} (\gamma_5\gamma^{\mu})^{(1)} (\gamma_5\gamma_{\mu})^{(2)}\nonumber\\
           &+\left({{\cal F}_{SO}(s,t,v)}1^{(1)}  1^{(2)}+ {{\cal F}_{VO}(s,t,v)} \gamma^{\mu(1)} \gamma_{\mu}^{(2)}
                      + {{\cal F}_{T0}(s,t,v)} \sigma^{\mu \nu (1)} \sigma_{\mu \nu}^{(2)}\right. \nonumber \\
                      &\left.- {{\cal F}_{PO}(s,t,v)} (i\gamma_5)^{(1)}  (i\gamma_5)^{(2)}
                      + {{\cal F}_{A0}(s,t,v)} (\gamma_5\gamma^{\mu})^{(1)} (\gamma_5\gamma_{\mu})^{(2)}\right)\frac{S^{(1)-1}(p_1)}{2m}\label{eq:FermiOffshell}
\end{align}
where
\begin{equation}
S^{(1)-1}(p_1)=\gamma^{(1)}\cdot p_1-m
\end{equation}
is the inverse of the propagator with four-momentum $p_1$.

The effect of this on a positive-energy spinor is
\begin{align}
\frac{S^{(1)-1}(p_1)}{2m}u(\bm{p}_1,\lambda_1)&=\frac{\gamma^{(1)0}p_1^0-\bm{\gamma}^{(1)}\cdot\bm{p}_1-m}{2m}u(\bm{p}_1,\lambda_1)\nonumber\\
&=\frac{\gamma^{(1)0}p_1^0-\gamma^{(1)0}E_{p_1}+\gamma^{(1)0}E_{p_1}-\bm{\gamma}^{(1)}\cdot\bm{p}_1-m}{2m}u(\bm{p}_1,\lambda_1)\nonumber\\
&=\left(\frac{\gamma^{(1)0}(p_1^0-E_{p_1})}{2m}-\Lambda^{(1)-}(p_1)\right)u(\bm{p}_1,\lambda_1)\nonumber\\
&=\frac{\gamma^{(1)0}(p_1^0-E_{p_1})}{2m}u(\bm{p}_1,\lambda_1)=\frac{v}{2m\sqrt{t}}\gamma^{(1)0}u(\bm{p}_1,\lambda_1),
\end{align}
where $E_{p_1} = E_t$ is the on shell energy of the spinor.
Since this is linear in $v$, the five off-shell terms must vanish on shell as expected. In addition, while the helicity matrix elements needed to obtain the on-shell terms are such that $\lambda_1=\lambda'_1$ and $\lambda_2=\lambda'_2$, or $\lambda_1=-\lambda'_1$ and $\lambda_2=-\lambda'_2$, the off-shell terms require matrix elements where $\lambda_1=-\lambda'_1$ and $\lambda_2=\lambda'_2$, or $\lambda_1=\lambda'_1$ and $\lambda_2=-\lambda'_2$ due to the extra factor of $\gamma^{(1)0}$. Therefore, it is impossible to obtain the off-shell terms from the on-shell data. Some dynamical model of the interaction is required to determine these contributions. Therefore, for this work we will take
\begin{equation}
{\cal F}_{SO}= {\cal F}_{VO}={\cal F}_{T0}={\cal F}_{PO}={\cal F}_{A0}=0
\end{equation}
in (\ref{eq:FermiOffshell}).

Additional problems occur in analytic continuation of the Fermi invariants from the $t$ cm frame where $t\ge 4m^2$ and $s\le 0$ to the $s$ cm frame where $t\le 0$ and $s\ge 4m^2$.

\begin{figure}
    \centerline{\includegraphics[height=3in]{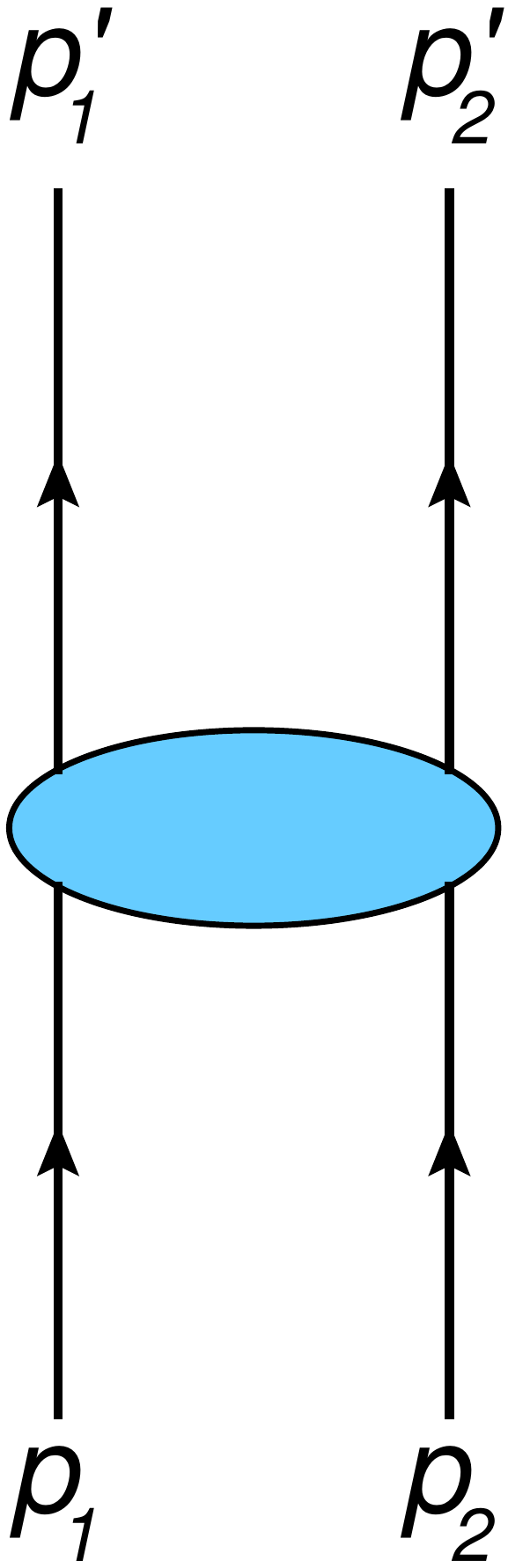}}
    \caption{(color online) Diagram representing $NN$ scattering in the $s$ channel cm frame. }
    \label{fig:offshellamp}
\end{figure}

\begin{figure}
    \centerline{\includegraphics[height=3in]{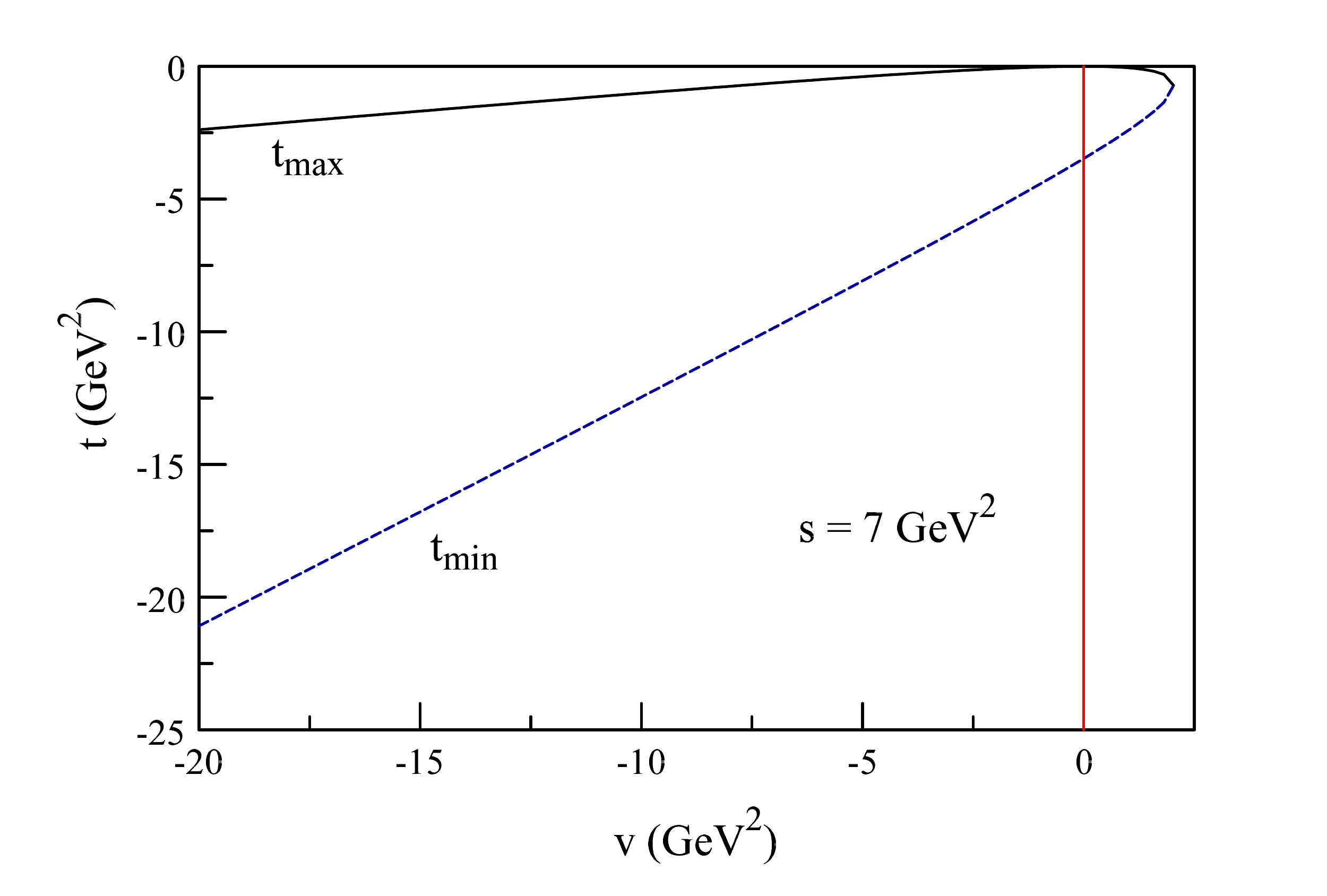}}
    \caption{(color online) The area between the curve labeled $t_{max}$ and $t_{min}$ contains the allowed values of $t$ in the $s$ cm frame as a function of the off-shell parameter $v$ for $s=7\ {\rm GeV^2}$. The line at $v=0$ shows the range of $t$ for on-shell scattering.} 
    \label{fig:trange}
\end{figure}

The scattering amplitude in the $s$-channel cm frame is represented by Fig. {\ref{fig:offshellamp}}
where
\begin{align}
p_1&=(p_1^0,\bm{p})\nonumber\\
p'_1&=(E,\bm{p'})\nonumber\\
p_2&=(E',-\bm{p})\nonumber\\
p'_2&=(E',-\bm{p}')\,.
\end{align}
As in the previous case $E'$ and $E$ represent on-shell energies and $p_1^0$ is the off-shell energy of particle 1 in the initial state. From conservation of energy
\begin{align}
E'&=\frac{\sqrt{s}}{2}\nonumber\\
|\bm{p}'|&=\sqrt{\frac{s}{4}-m^2}\nonumber\\
E&=\frac{s-v}{2\sqrt{s}}\nonumber\\
|\bm{p}|&=\sqrt{\frac{(s-v)^2}{4s}-m^2}\,.\label{eq:scmoff}
\end{align}
In this frame $t$ is given by
\begin{equation}
t=\frac{4m^2-s+v+\sqrt{(s-4m^2)\left(\frac{(s-v)^2}{s}-4m^2\right)}\cos\theta_{cm}}{2}
\end{equation}
where $\theta_{cm}$ is the scattering amplitude in the $s$ cm frame. As a result, $t$ is bounded by the functions
\begin{equation}
t_{max}=\frac{4m^2-s+v+\sqrt{(s-4m^2)\left(\frac{(s-v)^2}{s}-4m^2\right)}}{2}
\end{equation}
and
\begin{equation}
t_{min}=\frac{4m^2-s+v-\sqrt{(s-4m^2)\left(\frac{(s-v)^2}{s}-4m^2\right)}}{2}\,.
\end{equation}
The maximum value of $v$ is given by 
\begin{equation}
v_{max}=s-2m\sqrt{s}\,.
\end{equation}
These constraints are shown for $s=7\ {\rm GeV^2}$ in Fig. \ref{fig:trange}, where the physically accessible region is the area between the lines for $t_{max}$ and $t_{min}$.  
Note that the on-shell amplitudes are given by $v=0$ where $t_{max}=0$ and $t_{min}=4m^2-s$.

Now consider (\ref{eq:zoff}) which gives the value of the $t$ cm scattering angle for particle 1 off-shell. Evaluation of $z$ at the extreme values of $t$ in the $s$ cm frame gives
\begin{equation}
z(t_{max})=z(t_{min})=1
\end{equation}
for $v\neq 0$. However, for $v=0$,
\begin{equation}
z(t_{max})=\frac{s}{2m^2}-1\,.
\end{equation}
So $z$ as described by (\ref{eq:zoff}) is discontinuous at $v=0$. This is a problem for our Regge model parameterization of the scattering amplitudes since the Regge amplitudes are represented as
\begin{equation}
 R^{IPG}_{\pm j}(s,t) \propto \sum_{k}\xi_{k\pm}(t)\beta_{\pm k}^{IPG}(t) z^{\alpha_k(t)},\label{eq:ReggeExchange}
\end{equation}
where $\beta^{IPG}(t)\propto e^{\beta_1t}$ is the residue, $\xi_{\pm}(t)$ is a phase function and $\alpha(t)=\alpha_0+\alpha_1 t$ is the Regge trajectory. The discontinuity in $z$ results in very extreme discontinuous behavior in Regge amplitudes due to the factor of $z^{\alpha(t)}$. This is contrary to the reasonable expectation that the scattering amplitudes should have a smooth continuous extrapolation off-shell. Therefore, a straightforward analytic continuation of the off-shell Fermi invariants for the Regge model from the $t$ cm frame to the $s$ cm frame is unsatisfactory and an alternate method must be considered.

An alternate approach is to make the off-shell extrapolation after the on-shell Fermi invariants are analytically continued to the $s$ cm frame. This requires that we choose a prescription for $z$ off-shell that will be continuous in $v$. We choose 
\begin{equation}
z=\frac{s(t+t_{max}-2t_{min})-4m^2(t-t_{min})}{s(t_{max}-t)+4m^2(t-t_{min})}
\end{equation}
which is constrained such that
\begin{equation}
z(t_{max})=\frac{s}{2m^2}-1
\end{equation}
and
\begin{equation}
z(t_{min})=1
\end{equation}
for all allowed values of $v$.

The remainder of the details of the parameterization follow those in \cite{FVO_Reggemodel} and \cite{FJVO} with the exception that Eqn. (19) of \cite{FJVO} is replaced by
\begin{align}
\Xi_{S1}(s,t,v)&=-\frac{m^2}{2(4m^2-t)}\nonumber  \\
\Xi_{V2}(s,t,v)&=-\frac{4m^2-t}{8\left(2s+t-4m^2-v \right)}\nonumber  \\
\Xi_{V3}(s,t,v)&=\frac{t}{8(2s+t-4m^2-v)}\nonumber  \\
\Xi_{T3}(s,t,v)&=-\frac{m^2}{4(2s+t-4m^2-v)}\nonumber  \\
\Xi_{P4}(s,t,v)&=-\frac{m^2}{2t}\nonumber  \\
\Xi_{A5}(s,t,v)&=\frac{1}{8}\,.
\end{align}

\subsection{The off-shell cross section}

One way to visualize the nature of this off-shell prescription is the calculation of the $NN$ cross section off-shell. The helicity matrix elements of (\ref{eq:FermiOffshell}) are defined as
\begin{equation}
M_{\lambda'_1,\lambda'_2;\lambda_1,\lambda_2}(s,t,v)=\bar{u}^{(1)}(\bm{p}'_1,\lambda'_1)\bar{u}^{(2)}(\bm{p}'_2,\lambda'_2)\hat{M}(s,t,v)u^{(1)}(\bm{p}_1,\lambda_1)u^{(2)}(\bm{p}_2,\lambda_2)\,,
\end{equation}
where $u(\bm{p},\lambda)$ is a spinor in the helicity basis.
For identical particles, only five amplitudes are independent and are given by
\begin{align} \label{eq:amplitudes(abcde)}
a(s,t,v) &= \phi_1(s,t,v) = M_{\frac{1}{2},\frac{1}{2};\frac{1}{2},\frac{1}{2}}(s,t,v) \nonumber\\
b(s,t,v) &= \phi_5(s,t,v) = M_{\frac{1}{2},\frac{1}{2};\frac{1}{2},-\frac{1}{2}}(s,t,v)\nonumber\\
c(s,t,v) &= \phi_3(s,t,v) =  M_{\frac{1}{2},-\frac{1}{2};\frac{1}{2},-\frac{1}{2}}(s,t,v)\nonumber\\
d(s,t,v) &= \phi_2(s,t,v) = M_{\frac{1}{2},\frac{1}{2};-\frac{1}{2},-\frac{1}{2}}(s,t,v) \nonumber\\
e(s,t,v) &= \phi_4(s,t,v) =  M_{\frac{1}{2},-\frac{1}{2};-\frac{1}{2},\frac{1}{2}}(s,t,v)\,,
\end{align}
A fictitious off-shell cm differential cross section can then be defined as
\begin{equation}
\frac{d\sigma}{d\Omega_{cm}}(s,t,v)=\frac{m^4}{8\pi^2s}\left(|a|^2+4|b|^2+|c|^2+|d|^2+|e|^2\right)\,.
\end{equation}

\begin{figure}
    \centerline{\includegraphics[height=3in]{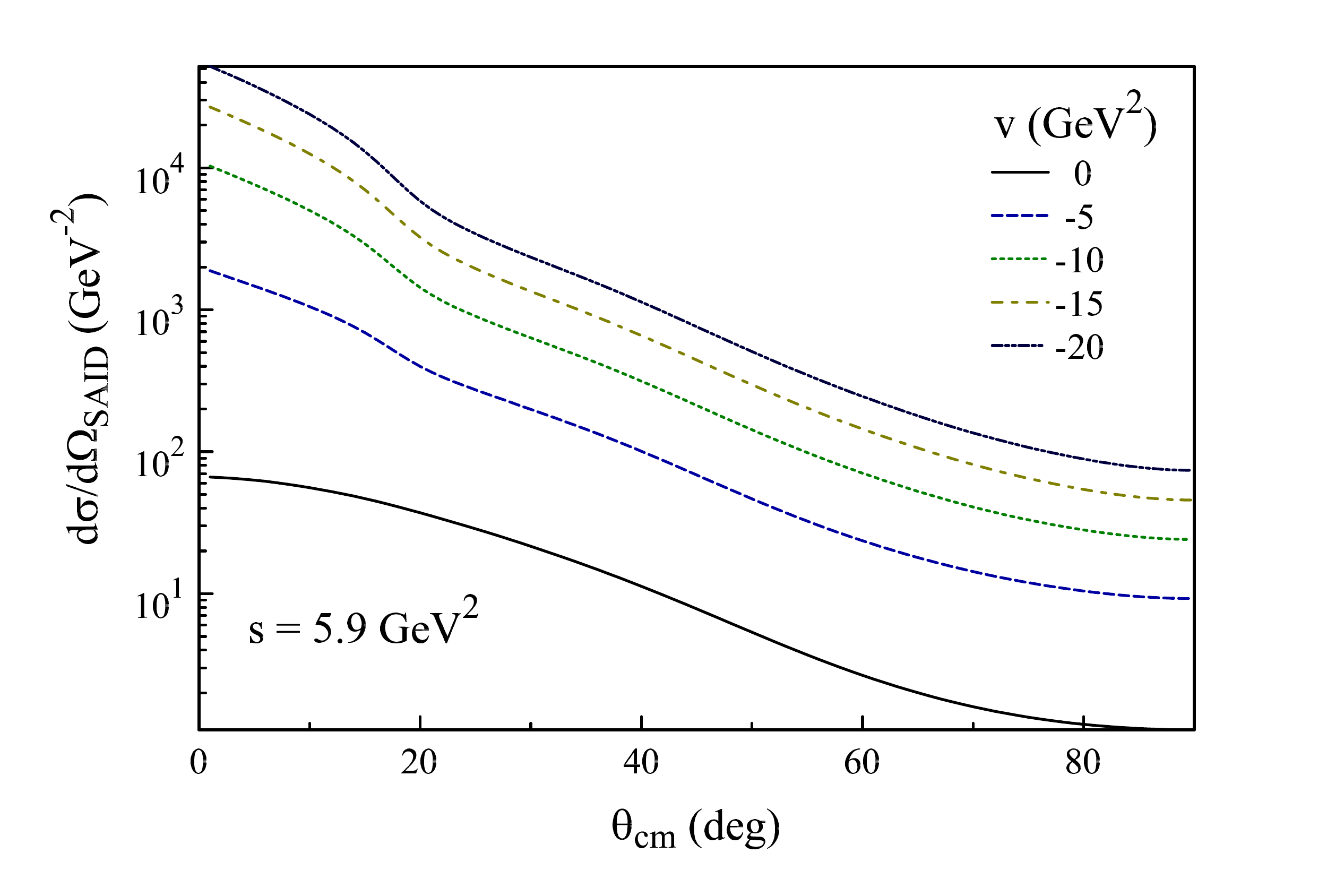}}
    \caption{(color online) Off-shell cross sections for $pp$ scattering using the SAID helicity amplitudes as a function of $v$ for $s=5.9\ {\rm GeV^2}$. }
    \label{fig:ppSAID}
\end{figure}

One source of $v$ dependence in the helicity amplitude results from the matrix elements of the dirac gamma-matrices in (\ref{eq:FermiOffshell}) using the definitions of the on-shell energies given by (\ref{eq:scmoff}). In \cite{JVO_2008_newcalc}, where the SAID helicity amplitudes were used to describe the FSI, an off-shell prescription was proposed that effectively only contains these contributions. The off-shell cm cross sections using the SAID amplitudes for $s=5.9\ {\rm GeV^2}$ is shown in Fig \ref{fig:ppSAID} for $pp$ scattering at various values of $v$.
Here $v=0$ corresponds to the physical cross section. Note that as the magnitude of $v$ increases the size of the cross section also increases. It can be shown by explicit calculation of the helicity matrix elements using this prescription that the amplitudes must vary as $|v|$ when the magnitude of $v$ becomes large. The cross sections should then vary as $v^2$. Clearly, use of this prescription in calculations of deuteron electrodisintegration will diverge unless a cutoff is introduced. In \cite{JVO_2008_newcalc}, we introduced a cutoff of the form
\begin{equation}
f(v)=\frac{(\Lambda^2-m^2)^2}{(\Lambda^2-m^2)^2+v^2}\label{eq:cutoff}
\end{equation}
where the cutoff mass $\Lambda$ was typically taken to be 1 GeV.

\begin{figure}
    \centerline{\includegraphics[height=3in]{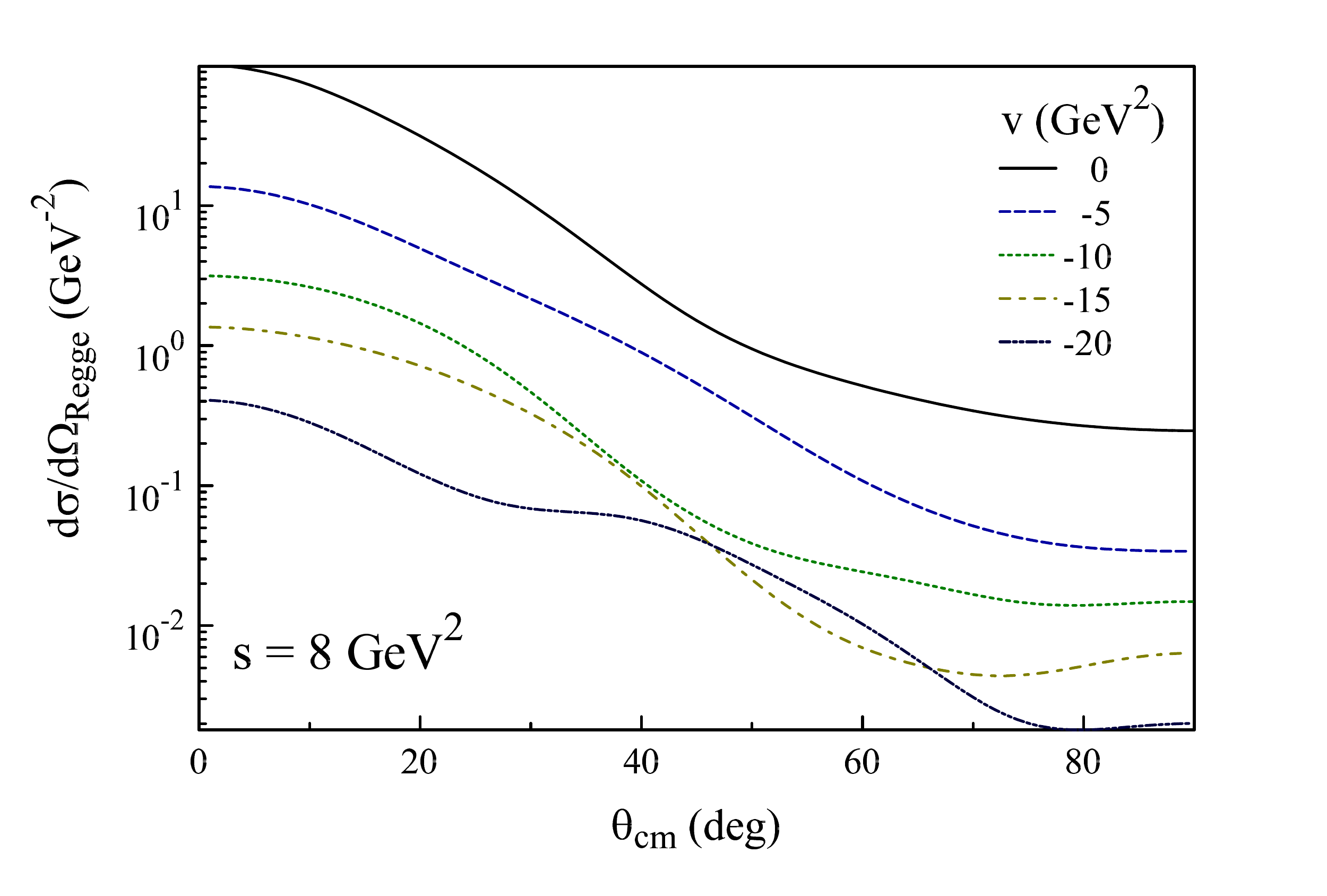}}
    \caption{(color online) Off-shell cross sections for $pp$ scattering using the Regge amplitudes as a function of $v$ for $s=8\ {\rm GeV^2}$.}  \label{fig:ppRegge}
\end{figure}

For the Regge model, the amplitudes (\ref{eq:ReggeExchange}) depend explicitly on $t$ through the phase factor $\xi(t)$, the Regge trajectory $\alpha(t)$ and the residue factor $\beta^{IPG}(t)\propto e^{\beta_1t}$. From Fig. \ref{fig:trange} it can be seen that the maximum value of $t$ becomes increasingly negative as $v$ moves away from the on-shell point. This means that the maximum size of the residue factor decreases exponentially away from the on-shell point. In addition, the range of $t$ values that can contribute increases as $v$ becomes more negative. Since the point where the  maximum value of $u$ occurs corresponds to the minimum value of $t$, this causes the overlap of the $t$ and $u$ channel contributions to decrease. We should therefore expect that the off-shell cross sections using the Regge model amplitudes should decrease exponentially as the magnitude of $v$ increases.
Figure \ref{fig:ppRegge} shows the off-shell $pp$ cross section at $s=8\ {\rm Gev^2}$ for various values of $v$.
This figure shows that the variation of the off-shell cross section with $v$ is consistent with the arguments made above, and decrease exponentially with increasing magnitude of $v$.

\section{Deuteron electrodisintegration}

\begin{figure}
    \centerline{\includegraphics[height=3in]{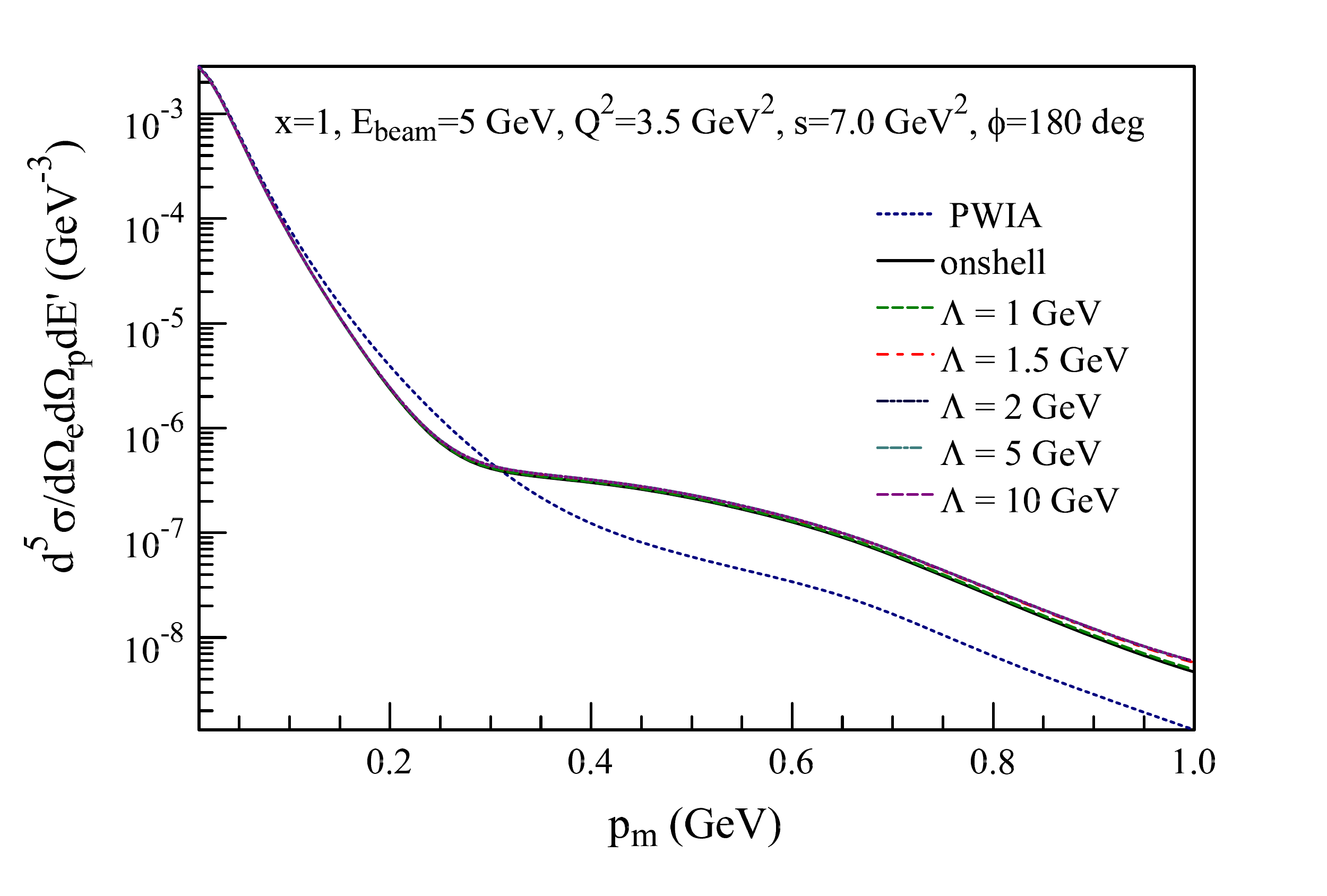}}
    \caption{(color online) The deuteron electrodisintegration cross section using the Regge FSI for $x=1$, beam energy $E_{beam}=5\ {\rm GeV}$, $Q^2=3.5\ {\rm GeV^2}$, $s=7.0\ {\rm GeV^2}$ and $\phi=180^\circ$.}
    \label{fig:deep1}
\end{figure}

We now consider the effects of including the off-shell contributions from Fig. \ref{fig:deepiaoff}(b) using the off-shell prescription for the Regge FSI described above. It is useful to include the cutoff (\ref{eq:cutoff}) as a means of studying the off-shell contributions. Figure \ref{fig:deep1} shows the differential cross section for $x=1$, beam energy $E_{beam}=5\ {\rm GeV}$, $Q^2=3.5\ {\rm GeV^2}$, $s=7.0\ {\rm GeV^2}$ and $\phi=180^\circ$. Calculations of the PWIA represented by Fig. \ref{fig:deepia}(a), the PWIA plus the on-shell contribution represented by Fig. \ref{fig:deepiaoff}(a), and the cross section for the PWIA, on-shell and off-shell contributions of Fig. \ref{fig:deepiaoff}(b) for various values of the cutoff mass $\Lambda$.
It is clear from this figure that at the chosen kinematics the off-shell effects are small. However, since this is a semi-log plot the relative size of the off-shell contributions can be better understood by considering the ratio off-shell to on-shell cross sections for various values of $\Lambda$ defined by
\begin{equation}
\sigma_R=\frac{\left(\frac{d^5\sigma}{d\Omega_ed\Omega_pdE'}\right)_\Lambda}{\left(\frac{d^5\sigma}{d\Omega_ed\Omega_pdE'}\right)_{\rm onshell}}\,.
\end{equation}
\begin{figure}
    \centerline{\includegraphics[height=3in]{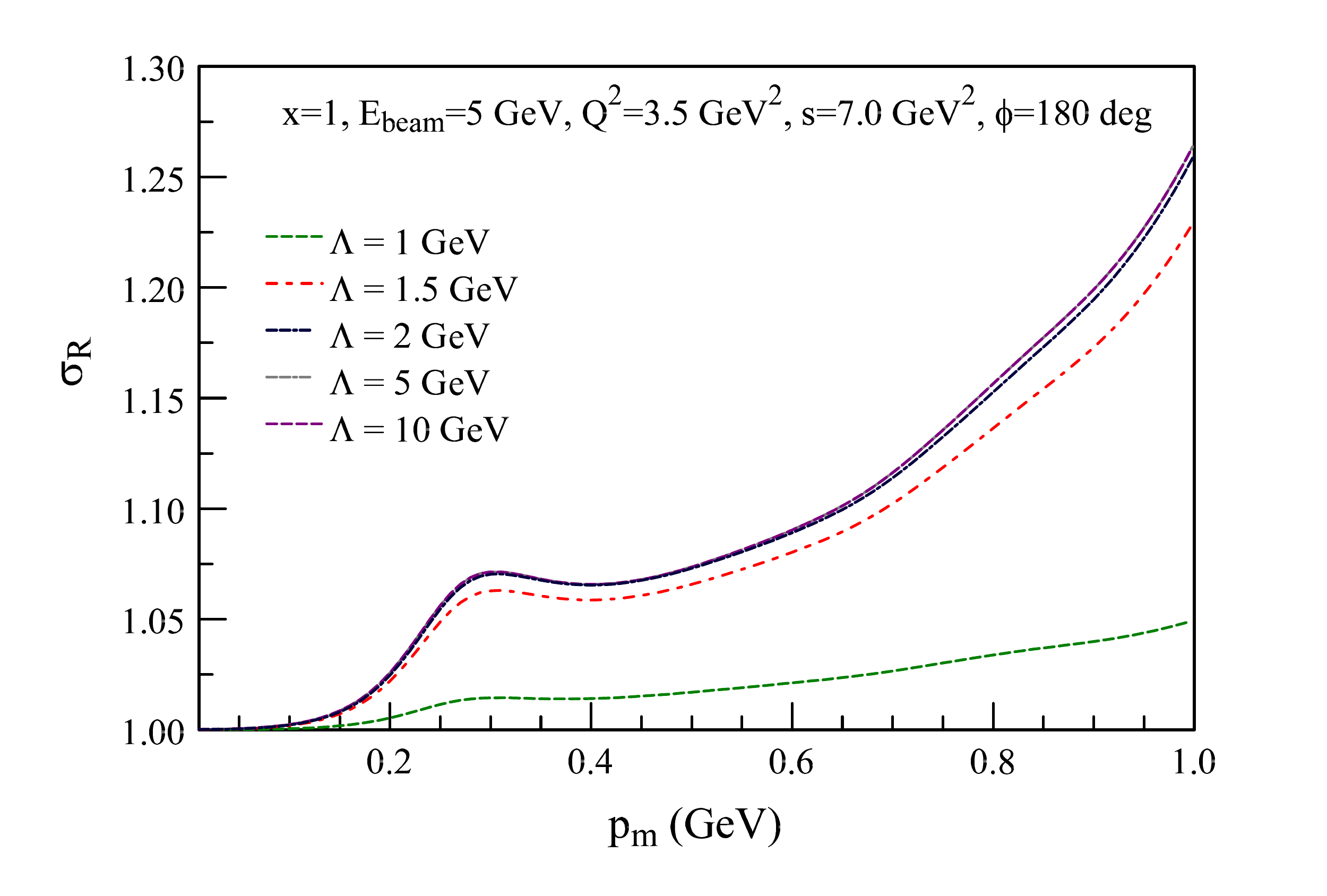}}
    \caption{(color online) Ratio of off-shell to on-shell cross sections for various values of the cutoff mass.}
    \label{fig:ratio1}
\end{figure}
This ratio is shown in Fig. \ref{fig:ratio1} for cutoff masses of 1 to 10 GeV. This shows that the off-shell contribution is quite sensitive to the cutoff mass for values below 2 GeV, but quickly saturates for larger values of the cutoff mass. Indeed the effect is effectively saturated by $\Lambda=10\ {\rm GeV}$, and we will use this value of the cutoff mass to represent the maximum variation in the off-shell contributions in subsequent figures. Note that the size of the off-shell effects increases with increasing missing momentum, and has maximum value for these kinematics of about 25 percent at $p_m=1.0\ {\rm GeV}$.

\begin{figure}
    \centerline{\includegraphics[height=3in]{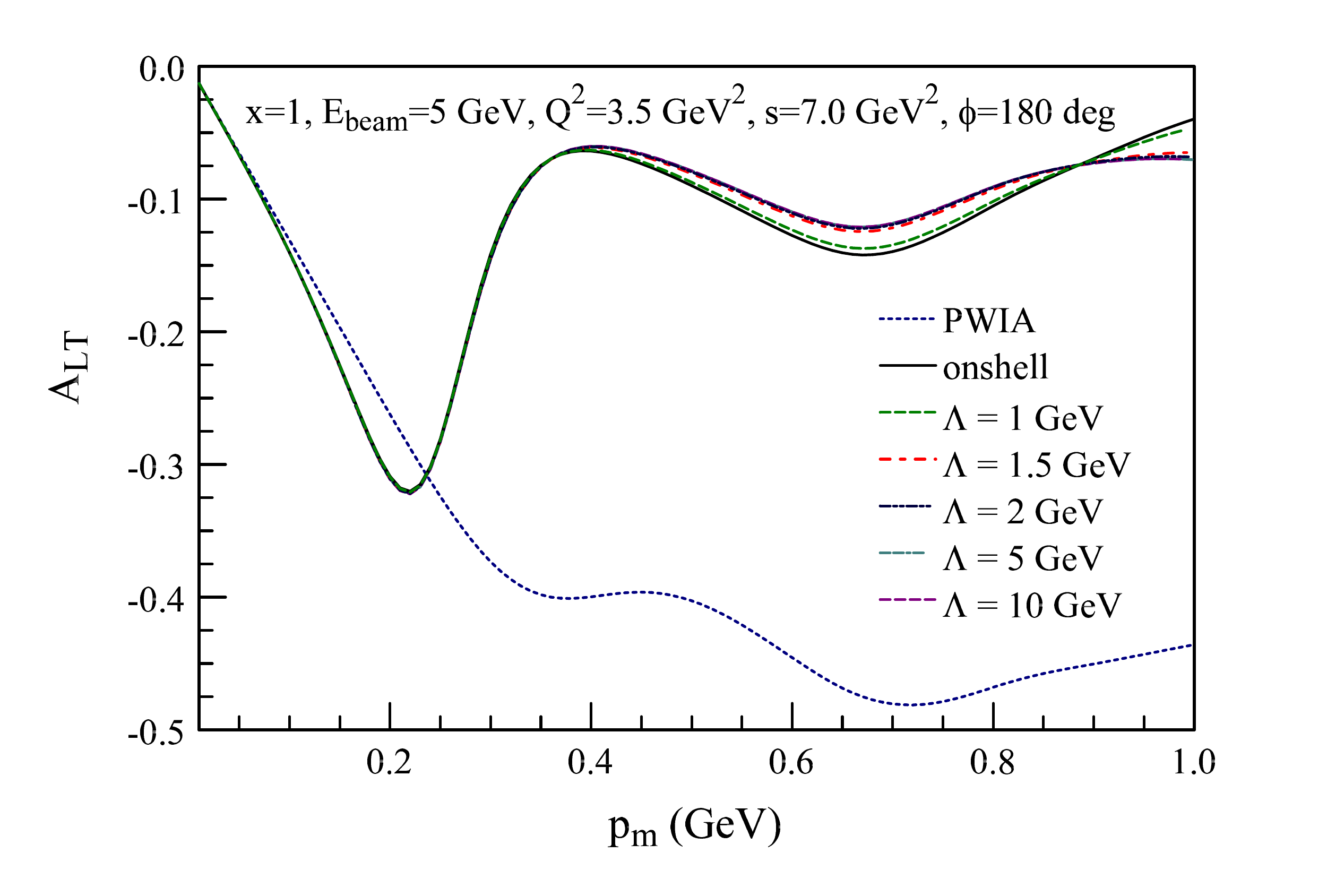}}
    \caption{(color online) The asymmetry $A_{LT}$ for PWIA, on-shell and off-shell FSI with different cutoff masses. }
    \label{fig:ALT}
\end{figure}

An example of off-shell contributions to the asymmetry $A_{LT}$ is shown in Fig. \ref{fig:ALT} for the same kinematics as in the previous figures. For this asymmetry the effect of including the on-shell FSI is large, but the contributions of off-shell scattering is small for these kinematics. As in the case of the cross sections, the rapid saturation of off-shell contributions is evident.

\begin{figure}
    \centerline{\includegraphics[height=3in]{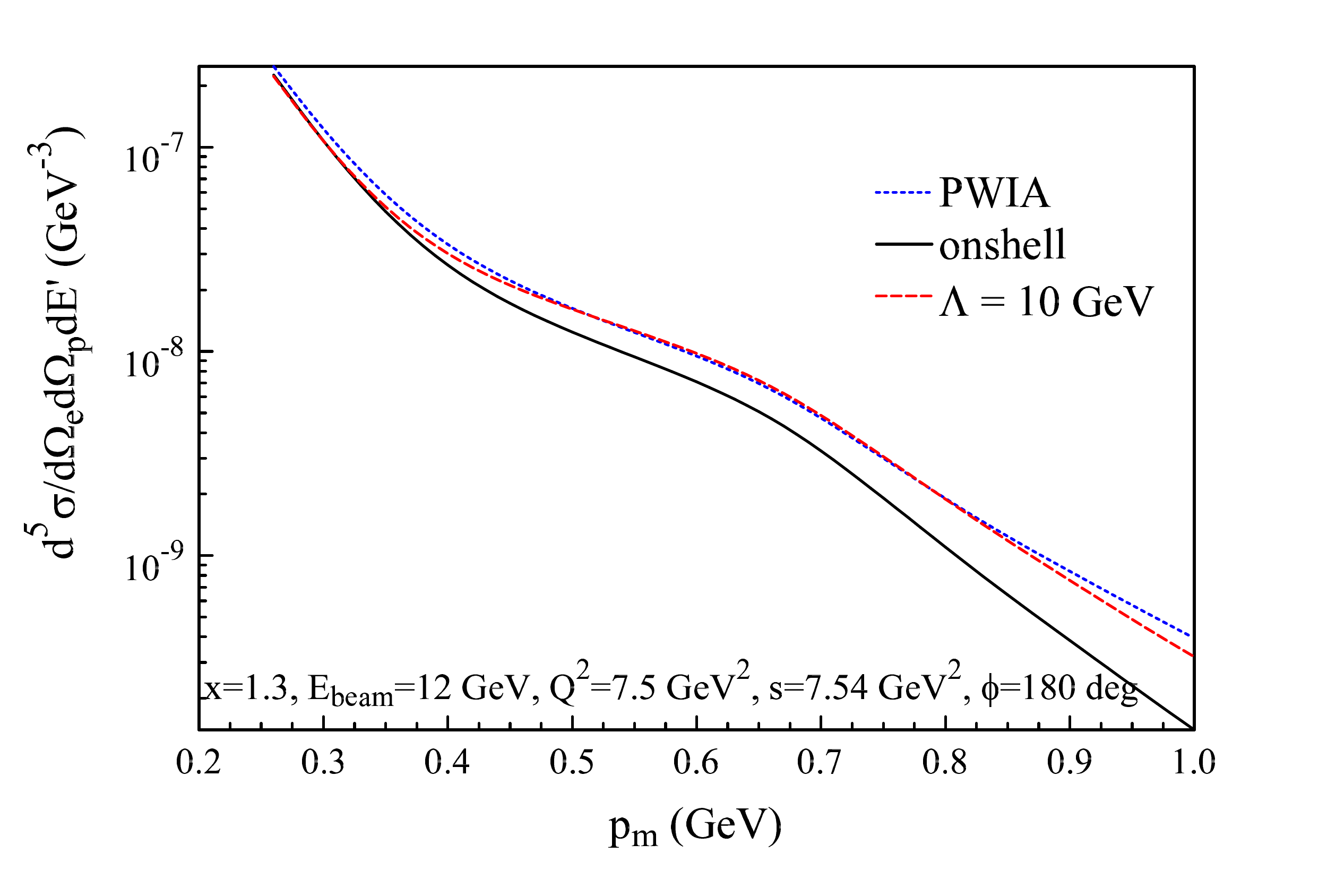}}
    \caption{(color online) Differential cross section for $x=1.3$, $E_{beam}=12\ {\rm GeV}$, $Q^2=7.5\ {\rm GeV^2}$, $s=7.54\ {\rm GeV^2}$ and $\phi=180^\circ$}.
    \label{fig:deep13}
\end{figure}

\begin{figure}
    \centerline{\includegraphics[height=3in]{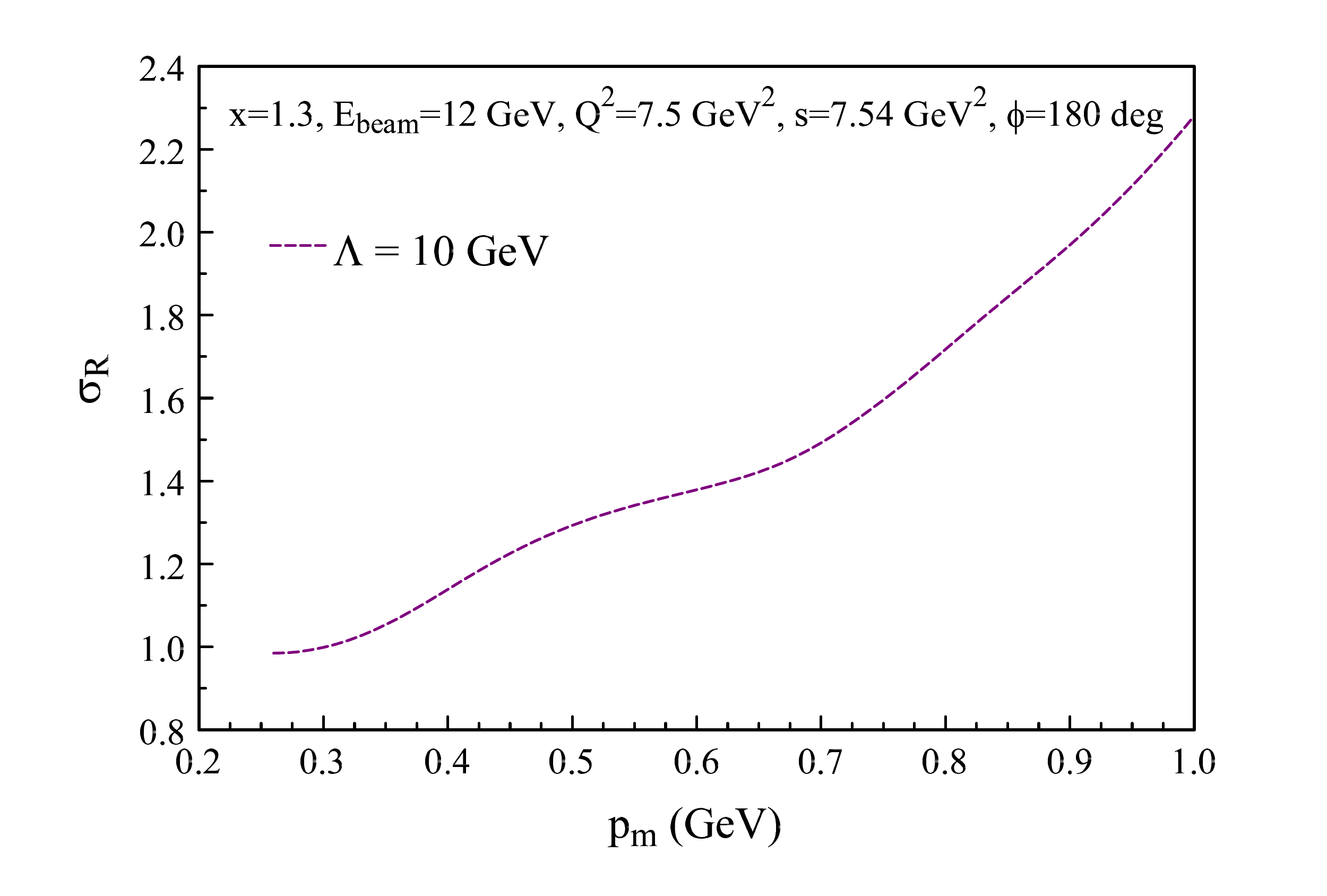}}
    \caption{(color online) Cross section ration $\sigma_R$ for the same kinematics as the previous figure. }
    \label{fig:ratio13}
\end{figure}

The differential cross section for $x=1.3$, $E_{beam}=12\ {\rm GeV}$, $Q^2=7.5\ {\rm GeV^2}$, $s=7.54\ {\rm GeV^2}$ and $\phi=180^\circ$ is shown in Fig. \ref{fig:deep13}. For these kinematics the on-shell cross section is substantial below the plane-wave cross section. For the $x=1$ kinematics of Fig. \ref{fig:deep1}, the on-shell cross section is larger than the PWIA cross section. In this case, we show only the case of saturated off-shell contributions with $\Lambda=10\ {\rm GeV}$. Here the off-shell contributions tend to move the cross section back toward the PWIA for most values of the missing momentum. 
The size of this effect can be seen by plotting the ratio of off-shell to on-shell cross sections as shown in Fig. \ref{fig:ratio13}. Clearly, for these kinematics the size of the off-shell contributions to the cross section are much larger than for the previous $x=1$ kinematics. The saturation value at $p_m=1\ {\rm GeV}$ is about 130 percent above the on-shell contribution as opposed to about 25 percent of the $x=1$ kinematics.

Since the off-shell prescription described here is somewhat arbitrary and incomplete, it is necessary to compare the computed cross sections to precision data to determine how large the effect should be. Unfortunately, there is a relatively small amount of data available in the kinematic regions where the Regge-model parameterization applies. The principal source of such data is from \cite{Boeglin_highmom}. The cross sections in this work are subdivided into a number of kinematical sets that display a large variation in the values of $x$ and $s$. In order to allow these to be compared for each of three values of $p_m$, the cross sections are normalized by the plane-wave calculation performed at each kinematic point. This is defined by the ratio
\begin{equation}
\sigma^{\rm PWIA}_R=\frac{\left(\frac{d^5\sigma}{d\Omega_ed\Omega_pdE'}\right)}{\left(\frac{d^5\sigma}{d\Omega_ed\Omega_pdE'}\right)_{\rm PWIA}}\,.
\end{equation}

\begin{figure}
    \centerline{\includegraphics[height=3in]{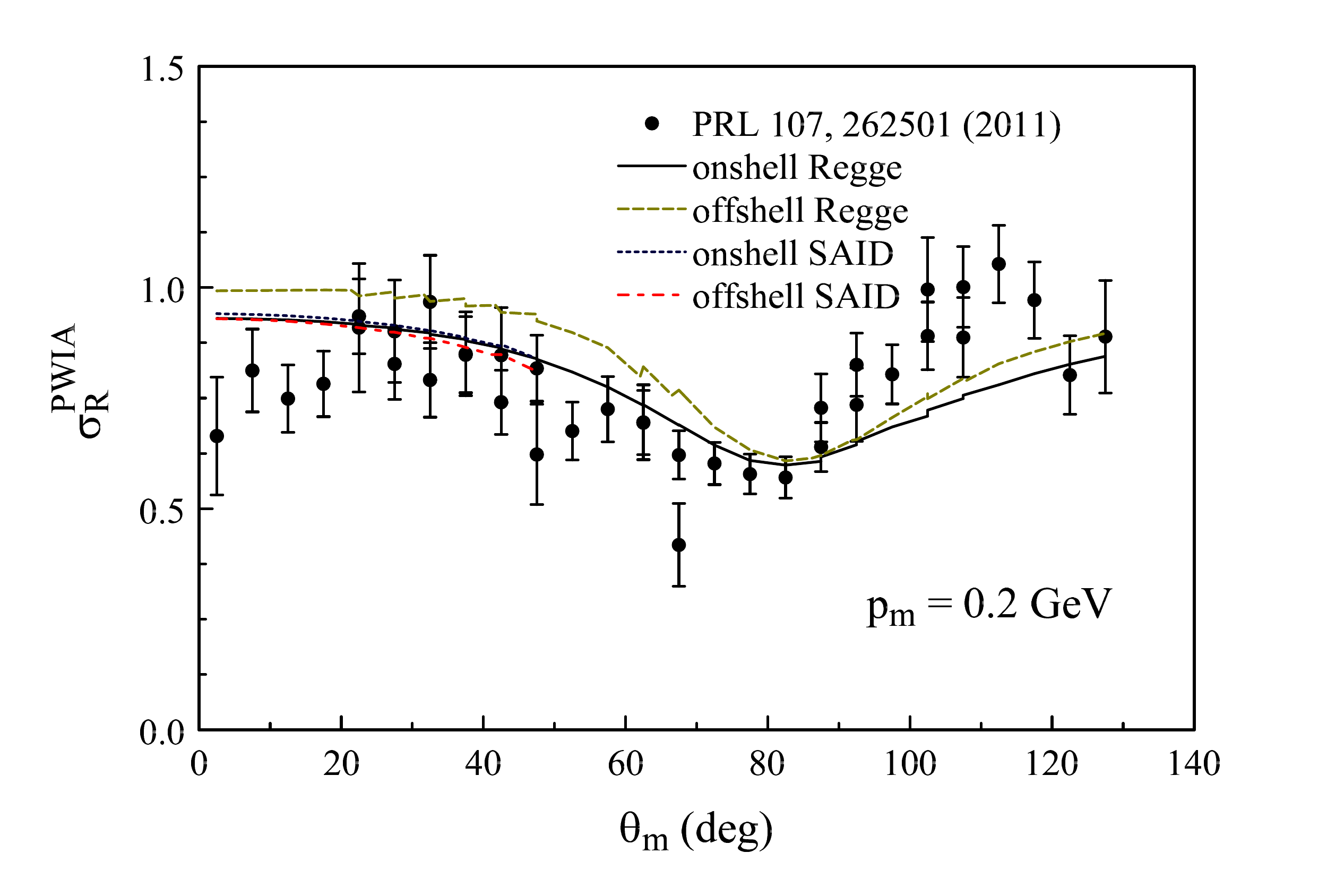}}
    \caption{(color online) Ratio of FSI cross sections to the PWIA cross section as a function of $\theta_m$ for $p_m=0.2\ {\rm GeV}$. Data are from \cite{Boeglin_highmom}. $\Lambda_{SAID} = 1\ {\rm GeV}$, and  $\Lambda_{Regge} = 10\ {\rm GeV}$.}
    \label{fig:werner2}
\end{figure}

\begin{figure}
    \centerline{\includegraphics[height=3in]{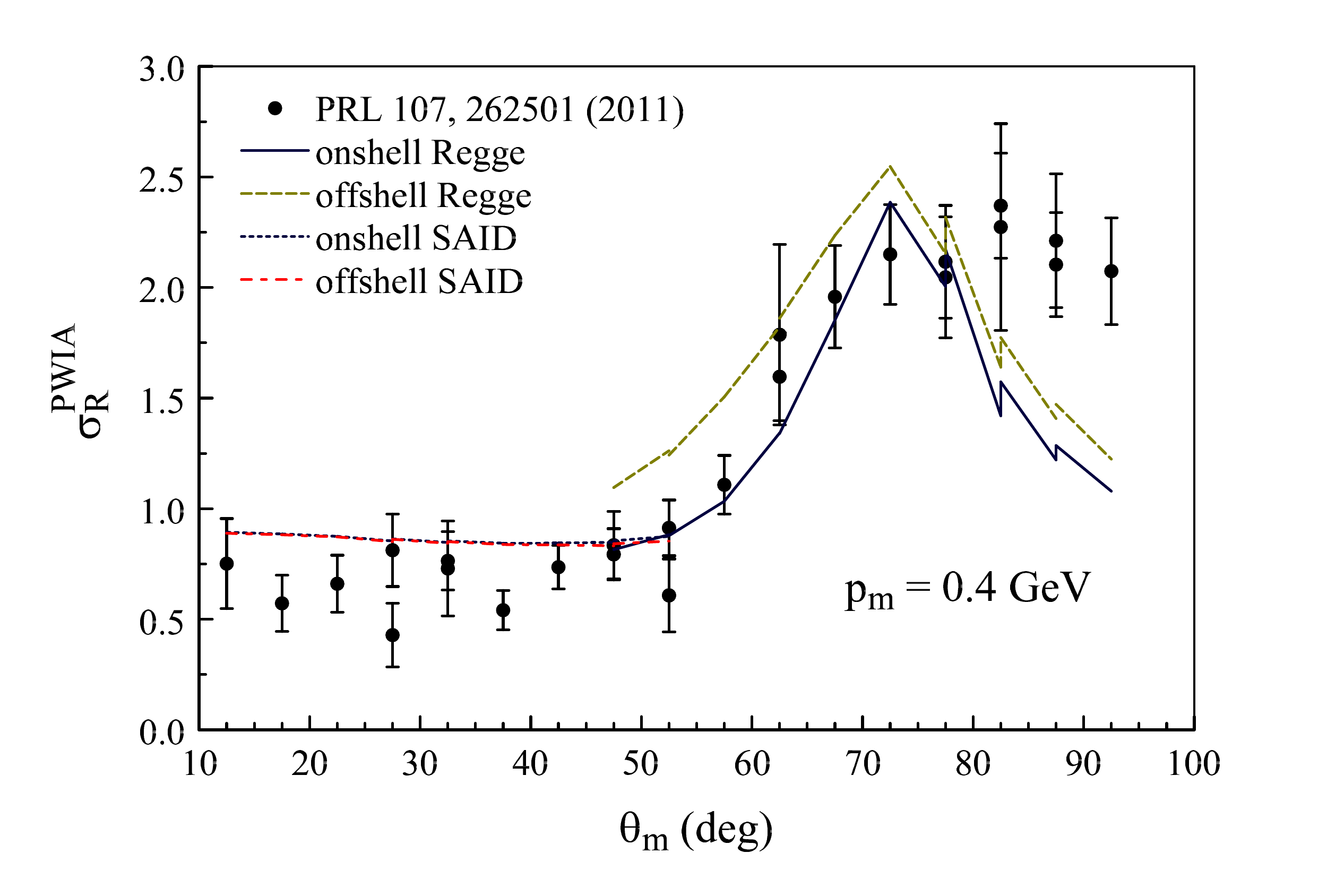}}
    \caption{(color online) Ratio of FSI cross sections to the PWIA cross section as a function of $\theta_m$ for $p_m=0.4\ {\rm GeV}$. Data are from \cite{Boeglin_highmom}. $\Lambda_{SAID} = 1\ {\rm GeV}$, and  $\Lambda_{Regge} = 10\ {\rm GeV}$.}
    \label{fig:werner4}
\end{figure}

\begin{figure}
    \centerline{\includegraphics[height=3in]{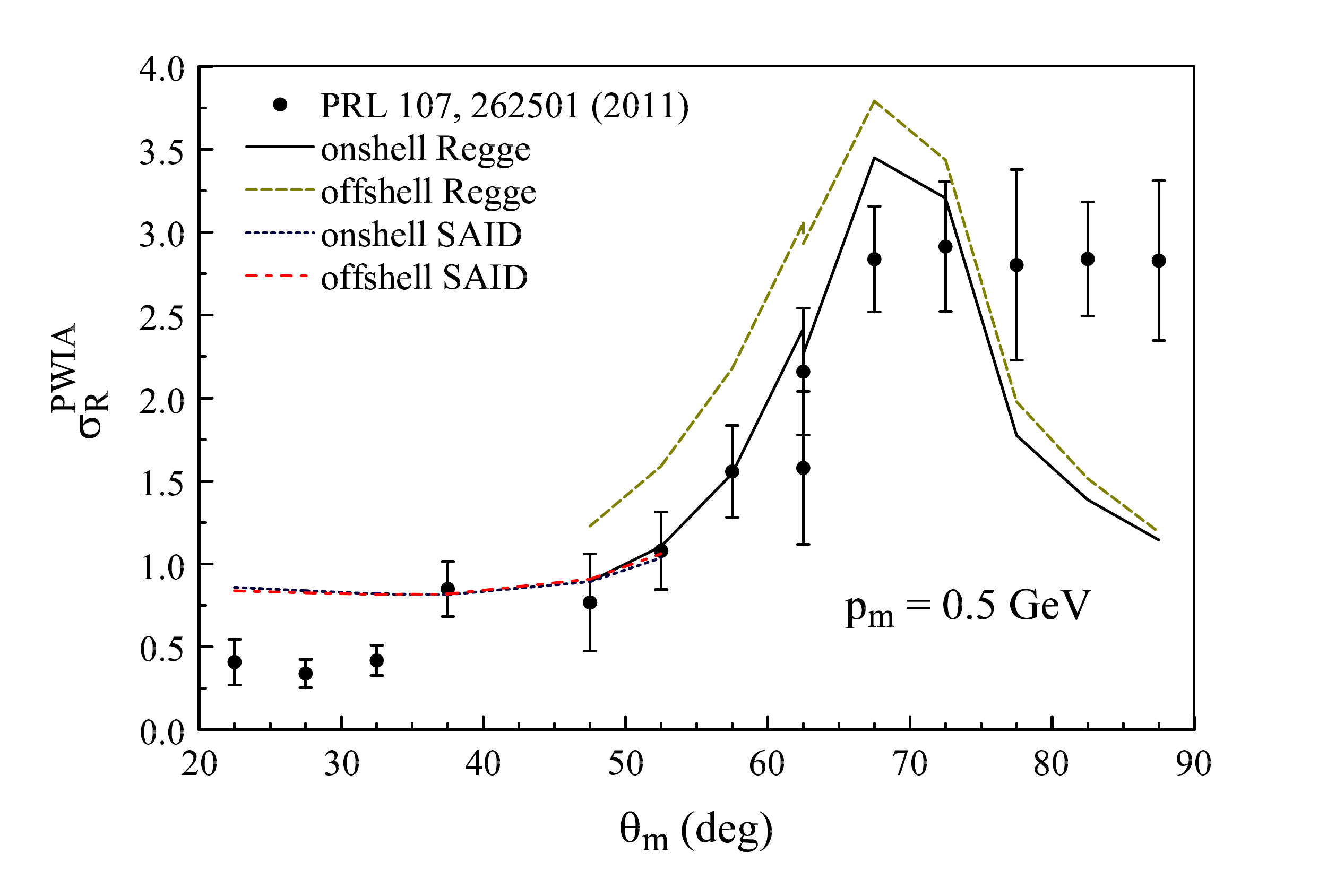}}
    \caption{(color online) Ratio of FSI cross sections to the PWIA cross section as a function of $\theta_m$ for $p_m=0.5\ {\rm GeV}$. Data are from \cite{Boeglin_highmom}.$\Lambda_{SAID} = 1\ {\rm GeV}$, and  $\Lambda_{Regge} = 10\ {\rm GeV}$.}
    \label{fig:werner5}
\end{figure}

The data and calculations for this quantity are shown in Fig. \ref{fig:werner2} for $p_m=0.2\ {\rm GeV}$, Fig. \ref{fig:werner4} for $p_m=0.4\ {\rm GeV}$ and Fig. \ref{fig:werner5} for $p_m=0.5\ {\rm GeV}$. For each case four calculations are shown. The calculation labeled ``onshell Regge'' uses the on-shell contribution from the Regge-model FSI. The calculation labeled ``offshell Regge'' uses the off-shell Regge-model FSI with $\Lambda=10\ {\rm GeV}$. The calculation labeled ``onshell SAID'' uses the on-shell SAID amplitudes and that labeled ``offshell SAID'' uses the off-shell prescription for the SAID amplitudes with $\Lambda=1\ {\rm GeV}$. In these figures smaller values of $\theta_m$ are associated with large values of $x$ and small values of $s$, and larger values of $\theta_m$ are associated with small values of $x$ and large values of $s$. Roughly, $0.8<x<1.5$ and $4.8\ {\rm GeV^2}<s<8.5\ {\rm GeV^2}$.
In all three cases all of the calculations agree qualitatively with shape of the data except at the largest and smallest angles. In regions where the SAID and Regge calculations overlap they are in close agreement. The calculations for the off-shell SAID FSI vary little from the corresponding on-shell results. It should be noted that none of the calculations shown in \cite{Boeglin_highmom} provides a satisfactory description of all of the data.

The calculations containing the off-shell Regge-model FSI along with the corresponding on-shell calculation provides the range of off-shell contributions that can be obtained with the prescription presented in this paper. For all three values of $p_m$ the off-shell calculation is larger than the on-shell calculation. Comparison with the data suggests that the off-shell contributions should be small. It would, however, be useful to use some caution in accepting this result until more data can be obtained.

\section{Summary and Conclusions}

In this work we have proposed a reasonable extrapolation of the Regge-model $NN$ scattering amplitude to the case where one of the initial nucleons is off-shell. This extrapolation is smooth and self regulating. Application of this approach to deuteron electrodistintegration show that the off-shell contributions are dependent on kinematics and are potentially large. However, comparison to the data from \cite{Boeglin_highmom} suggest that off-shell effects may be small. Caution should be taken in accepting this result until other data sets may become available. 

{\bf Acknowledgments}:   This work was
supported in part by funds provided by the U.S. Department of Energy
(DOE) under cooperative research agreement under No.
DE-AC05-84ER40150. The authors would like to thank Werner Boeglin for providing the data presented in this paper.
 
\bibliography{Regge_Deep}

\end{document}